\documentstyle[12pt]{article}
\input math_macros.tex

\def\ref#1{$^{#1)}$}
\def\ha{\widehat{C}}

\def\wt{\widetilde{C}}
\def\wwt{\widetilde{Q}}
\def\sl{\slash}
\begin{document}
\begin{titlepage}
\begin{center}
January 6, 1996     \hfill    LBL-38129 \\

\vskip .15in

{\large \bf Pole-Factorization Theorem in Quantum Electrodynamics}
\vskip .1in
Henry P. Stapp \\

{\em Lawrence Berkeley Laboratory\\
     University of California\\
     Berkeley, California 94720}

\end{center}

\vskip .05in

\begin{abstract}

In  quantum   electrodynamics a  classical  part of  the  S-matrix is  normally
factored  out in  order  to obtain  a  quantum  remainder  that can  be treated
perturbatively  without the occurrence  of infrared  divergences. However, this
separation, as  usually performed,  introduces spurious  large-distance effects
that produce  an apparent  breakdown  of the  important  correspondence between
stable particles and poles of the S-matrix, and, consequently, lead to apparent
violations  of  the   correspondence  principle and  to  incorrect  results for
computations in  the mesoscopic domain  lying between the  atomic and classical
regimes. An  improved  computational technique  is described  that allows valid
results  to be   obtained in  this  domain,  and that  leads,  for the  quantum
remainder, in the  cases studied, to  a physical-region  singularity structure
that, as   regards the  most   singular  parts,  is the  same  as the  normal
physical-region  analytic  structure  in theories  in which all  particles have
non-zero   mass. The  key  innovations  are to  define  the  classical  part in
coordinate   space,  rather  than in  momentum  space,  and to  define  there a
separation of the photon-electron coupling into its classical and quantum parts
that  has the   following  properties:  1) The   contributions  from  the terms
containing  only  classical  couplings can  be summed  to all  orders to give a
unitary  operator that  generates the  coherent  state that  corresponds to the
appropriate classical  process, and 2) The quantum  remainder can be rigorously
shown to  exhibit, as  regards  its most  singular  parts, the  normal analytic
structure. \end{abstract}

\vskip .2in
To appear in Annales de L'Institut Henri Poincare, 1996: Proceedings of 
Conference ``New Problems in the general theory of fields and particles''.

\end{titlepage}
\noindent{\bf 1. Introduction}

The pole-factorization property is the analog in quantum theory of the 
classical concept of the stable physical particle.
This property has been confirmed in a variety of rigorous
contexts$^{1,2,3}$ for theories in which the vacuum is the only state of zero 
mass. But calculations$^{4,5,6}$ have indicated that the property fails 
in quantum electrodynamics, due to complications associated with infrared 
divergences. Specifically, the singularity associated with the propagation of a
physical electron has been computed to be not a pole. Yet if the mass of the 
physical electron were $m$ and the dominant singularity of a scattering 
function at $p^2=m^2$ were not a pole then physical electrons would, according 
to theory, not propagate over laboratory distances like stable particles, 
contrary to the empirical evidence. 

This apparent difficulty with quantum
electrodynamics has been extensively studied$^{7,8,9}$, but not fully
clarified. It is shown here, at least in the context of a special case that is
treated in detail, that the apparent failure in quantum electrodynamics of 
the classical-type spacetime behaviour of electrons and positrons in the
macroscopic regime is due to 
approximations introduced to cope with infrared divergences.  Those divergences
are treated by factoring out a certain classical  
part, before treating the 
remaining part perturbatively. It can be shown, at least within the 
context of the case examined in detail, that if an accurate 
classical part of the photonic field is factored 
out then the required correspondence-principle and pole-factorization 
properties do hold.  The apparent failure of these latter two properties in 
references $4$ through $7$ are artifacts of 
approximations that are not justified in the context of the calculation 
of macroscopic spacetime properties: some factors $\exp ikx$ are replaced by
substitutes that introduce large errors for small $k$ but very large $x$.
 
The need to treat the factor $\exp ikx$ approximately arises from the fact
that the calculations are normally carried out in momentum space, where no 
variable $x$ occurs. The present approach is based on going to a mixed
representation in which both $x$ and $k$ appear. This is possible because the
variable $k$ refers to photonic degrees of freedom whereas the variable $x$
refers to electronic degrees of freedom.

To have a  mathematically well defined  starting point we  begin with processes
that have no charged  particles in the initial or  final states: the passage to
processes  where charged  particles are  present initially or  finally is to be
achieved by  exploiting the  pole-factorization property  that can be proved in
the simpler case considered first. To  make everything explicit we consider the
case where a  single charged particle  runs around a  spacetime closed loop: in
the Feynman  coordinate-space picture  the loop passes  through three spacetime
points, $x_1, x_2,$ and $x_3$, associated with, for example, an interaction with
a set of three  localized  external  disturbances. Eventually  there will be an
integration over  these variables. The  three regions are  to be far apart, and
situated  so  that a  triangular    electron/positron path   connecting them is
physically  possible.  To make  the  connection to  momentum  space, and to the
pole-factorization  theorem and correspondence  principle, we must study the
asymptotic behaviour of the amplitude as the three regions are moved apart.

Our procedure is based on the separation defined in reference 11  of the 
electromagnetic interaction operator into its ``classical'' and ``quantum'' 
parts. This separation is made in the following way. Suppose we first make a
conventional energy-momentum-space separation of the (real and virtual photons) into ``hard'' and
``soft'' photons, with hard and soft photons connected at ``hard'' and ``soft''
vertices, respectively. The soft photons can have small energies and momenta 
on the scale of the electron mass, but we shall not drop any ``small'' terms.
Suppose a charged-particle line runs from a hard
vertex $x^-$ to a hard vertex $x^+$. Let soft photon $j$ be coupled into this 
line at point $x_j$, and let the coordinate 
variable $x_j$ be converted by Fourier transformation to the associated 
momentum variable $k_j$.
Then the interaction operator $-ie\gamma_{\mu_j}$ is separated into its 
``classical'' and ``quantum'' parts by means of the formula
$$
-ie \gamma_{\mu_j}= C_{\mu_j} + Q_{\mu_j}, \eqno(1.1)
$$
where 
$$
C_{\mu_j} = -ie{z_{\mu_j}\over z\cdot k_{j}} \slash{k}_j, \eqno(1.2)
$$
and $z=x^+ - x^-$.

This separation of the interaction allows a corresponding separation of soft
photons into ``classical'' and ``quantum'' photons: a ``quantum'' photon has a
quantum coupling on at least one end; all other photons are called ``classical''
photons. 

The full contribution from all classical photons is represented in an
extremely neat and useful way. Specialized to our case of a single 
charged-particle loop $L(x_1, x_2, x_3)$ the key formula reads
$$
F_{op}(L(x_1,x_2, x_3))=:U(L(x_1,x_2,x_3)) F'_{op} (L(x_1,x_2,x_3)):.\eqno(1.3)
$$
Here $F_{op} (L(x_1, x_2, x_3))$ is the Feynman {\it operator} corresponding 
to the sum of contributions from {\it all} 
photons coupled into the charged-particle loop $L(x_1, x_2, x_3)$, and 
$F_{ op}'(L(x_1, x_2, x_3))$ is the analogous operator if
all contributions from classical photons are excluded.
The operators $F_{ op}$ and $F'_{ op}$ are both normal ordered operators: i.e., 
they are operators in the asymptotic-photon Hilbert space, and the destruction 
operators of the incoming photons stand to the right of the creation 
operators of outgoing photons. On the right-hand side of $(1.3)$ all of the 
contributions corresponding to classical photons are included in 
the unitary-operator factor $U(L)$ defined as follows:
$$
U(L) = e^{<a^*\cdot J(L)>} e^{-\half <J^*(L)\cdot J(L)>}
      e^{-<J^*(L)\cdot a>}e^{i\Phi (L)}. \eqno(1.4)
$$
Here, for any $a$ and $b$, the symbol $<a\cdot b >$ is an abbreviation for the integral
$$
<a\cdot b>\equiv \int {d^4k\over (2\pi)^4} 2\pi \theta (k_0)\delta (k^2) 
a_\mu(k)(-g^{\mu\nu} )b_\nu(k),\eqno(1.5)
$$
\noindent and $J(L,k)$ is formed by integrating $\exp ikx$ around the loop $L$: 
$$
J_\mu(L,k) \equiv \int_L dx_\mu e^{ikx}.\eqno(1.6)
$$
This classical current $J_{\mu}(L)$ is conserved:
$$
k^\mu J_\mu (L, k) =0. \eqno(1.7)
$$
The $a^*$ and $a$ in $(1.4)$ are photon creation and destruction operators, 
respectively, and
$\Phi (L)$ is the classical action associated with the motion of a charged
classical particle along the loop $L$:
$$
\Phi (L) = {(-ie)^2\over 8\pi} \int_L dx'_{\mu} g^{\mu\nu} \int_L dx''_\nu
\delta((x' - x'')^2)\eqno(1.8)
$$
The operator $ U(L)$ is {\it pseudo} unitary if it is written in explicitly 
covariant form, but it can be reduced to a strictly
unitary operator using by $(1.7)$ to eliminate all but the two
transverse components of $a_\mu (k),a^*_\mu (k), J_\mu(k)$, and $J^*_\mu (k)$.

The colons in (1.3) indicate that the creation-operator parts of the normal-
ordered operator $F'_{op}$ are to be placed on the left of $U(L)$.

The unitary operator $U(L)$ has the following property:
$$
U(L)|vac > = |C(L) >. \eqno(1.9)
$$
Here $|vac>$ is the photon vacuum, and $|C(L)>$ represents the normalized coherent state corresponding to 
the classical 
electromagnetic field radiated by a charged classical point particle
moving along the closed spacetime loop $L$,
in the Feynman sense.

The simplicity of (1.3) is worth emphasizing: it says that the complete effect 
of all classical photons is contained in a simple  unitary  operator   that is
independent of the quantum-photon contributions: this factor is a well-defined 
unitary operator that depends only on the (three)
hard vertices $x_1, x_2$, and $x_3$. It is independent of the remaining
details of $F'_{op}(L(x_1, x_2, c_3))$, even though the
classical couplings are originally interspersed in all possibly ways among the 
quantum couplings that appear in $F'_{op}(L(x_1, x_2, x_3))$.
The operator $U(L)$ supplies the classical bremsstrahlung-radiation photons 
associated with the deflections of the charged particles that occur at the
three vertices, $x_1,x_2,$ and $x_3$.

Block and Nordsieck$^{12}$ have already emphasized that the infrared
divergences arise from the classical aspects of the elecromagnetic field.
This classical component is exactly supplied by the factor $U(L)$.
One may therefore expect the remainder $F'_{op} (L(x_1, x_2, x_3))$ to be 
free of infrared problems: if we  
transform $F'_{op}(L(x_1,x_2,x_3))$ into momentum space, then it should satisfy
the usual pole-factorization property. A primary goal of this work is to show 
that this pole-factorization property indeed holds. To recover the physics one 
transforms $F'_{op}$ to coordinate space,
and then incorporates the real and virtual classical photons by using 
$1.3$ and $1.4$.

The plan of the paper is as follows.
In the following section  2 rules are established for writing down the
functions of interest directly in momentum space.
These rules are expressed in terms of operators that act on 
momentum--space Feynman functions and yield momentum--space functions, with
classical or quantum interactions inserted into the charged-particle lines
in any specified desired order.

It is advantageous always to sum together the contributions corresponding
to all ways in which a photon can couple with C--type coupling into each
individual side of the triangle graph $G$. This sum can be expressed as a 
sum of just two terms. In one term the photon is coupled at one endpoint,
$x^+$, of this side of $G$, and in the other term the photon is coupled into 
the other end point, $x^-$, of this side of $G$.
Thus all C--type couplings become converted into couplings at the hard--photon
vertices of the original graph $G$. 

This conversion introduces an important 
property. The charge--conservation (or gauge) condition $k^\mu J_\mu =0$ 
normally does not hold in quantum electrodynamics for individual graphs: 
one must sum over all ways in which the photon can be inserted into the graph.
But in the form we use, with each quantum vertex $Q$ coupled into the interior
of a line of $G$, but each classical vertex $C$ placed at a hard--photon 
vertex of $G$, the charge--conservation equation (gauge invariance) holds for 
each vertex separately: $k^\mu J_\mu =0$ for each vertex.

In section 3 the modification of the charged--particle propagator caused by 
inserting a single quantum vertex $Q_\mu$ into a charged-particle line is 
studied in detail. The resulting (double) propagator is re--expressed as 
a sum of three terms.
The first two are ``meromorphic'' terms having poles at $p^2=m^2$ and $p^2
= m^2-2pk -k^2$, respectively, in the variable $p^2$.
Because of the special form of the quantum coupling $Q_\mu$ each residue is of 
first order in $k$, relative to what would have been obtained with the usual 
coupling $\gamma_\mu$. This extra power of $k$ will lead to the infrared 
convergence of the residues of the pole singularities.

Our proof that this convergence property holds can be regarded as a 
systematization and confirmation of the 
argument for infrared convergence given by Grammer and Yennie$^{13}$.

The third term is a nonmeromorphic contribution.
It is a difference of two logarithms. This {\it difference} has a power of
$k$ that renders the contribution infrared finite.

\noindent
{\bf 2.  Basic Momentum--Space Formulas}

The separation of the soft--photon interaction into its quantum and 
classical parts is defined in Eq. (1.1).
This separation is defined in a mixed representation in which hard 
photons are represented in coordinate space and soft photons are 
represented in momentum space.
In this representation one can consider a ``generalized propagator''.
It propagates a charged particle from a hard--photon vertex $y$ to a 
hard--photon vertex $x$ with, however, the insertion of soft--photon 
interactions.

Suppose, for example, one inserts the interactions with two soft photons 
of momenta $k_1$ and $k_2$ and vector indices $\mu_1$ and $\mu_2$.
Then the generalized propagator is 
$$
\eqalignno{
P_{\mu_1, \mu_2} &(x,y; k_1, k_2)\cr
&= \int {d^4p\over (2\pi )^4} e^{-ipx + i(p+k_1+k_2)y}\cr
&\times {i\over \slash{p}-m+i0}\gamma_{\mu_1}{i\over 
\slash{p}+\slash{k}_1-m+i0}\gamma_{\mu_2}{i\over 
\slash{p}+\slash{k}_1+\slash{k}_2-m+i0}.&(2.1)\cr}
$$
The generalization of this formula to the case of an arbitrary number of 
inserted soft photons is straightforward.
The soft--photon interaction $\gamma_{\mu_j}$ is separated into its parts 
$Q_{\mu_j}$ and $C_{\mu_j}$ by means of (1.1), with the $x$ and $y$ defined 
as in (1.2).

This separation of the soft--photon interaction into its quantum and 
classical parts can be expressed also directly in momentum space.
Using (1.1) and (1.2), and the familiar identities
$$
{1\over \slash{p}-m} \slash{k}
{1\over \slash{p} + \slash{k} - m} = {1\over 
\slash{p}-m} - {1\over \slash{p}+ \slash{k}-m},\eqno(2.2)
$$
and
$$
\left( - {\partial\over \partial p^\mu}\right)
{1\over \slash{p}-m} = {1\over 
\slash{p}-m} \gamma_\mu {1\over \slash{p}-m},\eqno(2.3)
$$
one obtains for the (generalized) propagation from $y$ to $x$, with a 
single classical interaction inserted, the expression (with the symbol 
$m$ standing henceforth for $m-i0$) 
$$
\eqalignno{
P_\mu(x,y; C,k)
&= \int {d^4p\over (2\pi )^4} 
\left( 
{i\over \slash{p}-m}\slash{k}
{i\over \slash{p}+\slash{k}-m}\right) 
{z_\mu\over zk+io} e^{-ipz+iky}\cr
&= \int {d^4p \over (2\pi )^4} e^{-ipz+iky}
\int^1_0 
d\lambda\left(-i {\partial\over\partial p^\mu}\right)
\left({i\over \slash{p}+\lambda\!\slash{k}-m}
\right)\cr
&&(2.4)\cr}
$$

The derivation of this result is given in reference 14.
Comparison of the result (2.4) to (2.1) shows that the result in 
momentum space of inserting a single quantum vertex $j$ into a 
propagator $i(\slash{p}-m)^{-1}$ is produced by the action of the 
operator
$$
\widehat{C}_{\mu_j} (k_j)= \int^1_0 d\lambda_j O(p\to p+ 
\lambda_j k_j)\left(-i{\partial\over \partial p^{\mu_j}}\right)\eqno(2.5)
$$
upon the propagator $i(\slash{p}- m)^{-1}$ that was present 
{\it before} the insertion of the vertex $j$.
One must, of course, also increase by $k_j$ the momentum entering the 
vertex at $y$.
The operator $O(p\to p+\lambda_jk_j)$ replaces $p$ by $p+\lambda_jk_j$.

This result generalizes to an arbitrary number of inserted classical 
photons, and also to an arbitrary generalized propagator: the 
momentum--space result of inserting in all orders into any generalized 
propagator $P_{\mu_1, \cdots , \mu_n} 
(p; k_1, \cdots , k_n)$ a set of $N$ 
classically interacting photons with $j= n+1, \cdots, n+N$ is 

$$
\eqalignno{
&\prod^{n+N}_{j=n+1} 
\widehat{C}_{\mu_j}(k_j) 
P_{\mu_1, \cdots , \mu_n} 
(p; k_1, \cdots , k_n)       
=\int^1_0 \ldots \int^1_0 d\lambda_{n+1}\ldots d\lambda_{n+N} 
\prod^{N}_{j=1} \left( -i{\partial\over \partial p^{\mu_{n+j}}}\right)\cr
&\hbox{\hskip.25in} P_{\mu_1, \cdots, \mu_n} (p+a; k_1, \cdots , k_n)&(2.6)\cr}
$$
where $a= \lambda_{n+1} k_{n+1} + \cdots + \lambda_{n+N} k_{n+N}$.
The operations are commutative, and one can keep each $\lambda_j=0$ 
until the integration on $\lambda_j$ is performed.

One may not wish to combine the results of making insertions in all 
orders.
The result of inserting the classical interaction at just one 
place, identified by the subscript $j\epsilon\{1,\cdots ,n\}$, 
into a (generalized) 
propagator $P_{\mu_1 \cdots \mu_n} (p; k_1, \cdots , k_n )$,
abbreviated now by $P_{\mu_j}$, 
is produced by the action of 
$$
\eqalignno{
\widetilde{C}_{\mu_j}(k_j)&\equiv\cr
&\int^\infty_{0} d\lambda_j O(p_i\to p_i+ \lambda_jk_j 
)\left(-{\partial\over \partial p^{\mu_j}}\right)&(2.7)\cr}
$$
upon $k_j^{\sigma_j}P_{\sigma_j}$.

There is a form analogous to (2.7) for the Q interaction: 
the momentum--space result produced by the insertion of a Q 
coupling into $P_{\mu_1\cdots \mu_n}(p; k_1,\cdots k_\mu) = P_{\mu_j}$
at the vertex identified by $\mu_j$ 
is given by the action of 
$$
\widetilde{Q}_{\mu_j}(k_j) \equiv (\delta_{\mu_j}^{\sigma_j}k_j^{\rho_j} 
- \delta_{\mu_j}^{\rho_j} 
k_j^{\sigma_j})\widetilde{C}_{\rho_j}(k_j)\eqno(2.8)
$$
upon $P_{\sigma_j}$ .

An analogous operator can be applied for each quantum interaction.
Thus the generalized momentum--space propagator represented by a line $L$ 
of a graph $G$ 
into which $n$ quantum interactions are inserted in a fixed order is 
$$
\eqalignno{
&P_{\mu_1\cdots \mu_n}(p; Q, k_1,Q, k_2, \cdots Q, k_n)=\cr
&\prod^n_{j=1} \left[ \int^\infty_0 
d\lambda_j(\delta_{\mu_j}^{\sigma_j}k_j^{\rho_j}- 
\delta^{\rho_j}_{\mu_j} k_j^{\sigma_j})\left( 
-{\partial\over \partial p^{\rho_j}}\right)\right]\cr
&\Big({i\over 
\slash{p}+\slash{a}-m}\gamma_{\sigma_1}{i\over\slash{p}+\slash{a} + 
\slash{k}_1-m} \gamma_{\sigma_2}{i\over 
\slash{p}+\slash{a}+\slash{k}_1+\slash{k}_2-m}\cr
&\cdots \times \gamma_{\sigma_n}{i\over 
\slash{p}+\slash{a}+\slash{k}_1+\cdots \slash{k}_n - 
m}\Big),&(2.9)\cr}
$$
where
$$
a= \lambda_1 k_1 + \lambda_2 k_2 + \cdots \lambda_n k_n.\eqno(2.10)
$$

If some of the inserted interactions are classical interactions then the 
corresponding factors
$(\delta_{\mu_j}^{\sigma_j}k_j^{\rho_j} - 
\delta_{\mu_j}^{\rho_j}k_j^{\sigma_j})$ are replaced by 
$(\delta_{\mu_j}^{\rho_j}k_j^{\sigma_j})$.

These basic momentum--space formulas provide the starting point for our 
examination of the analyticity properties in momentum space,
and the closely related question of infrared convergence.

One point is worth mentioning here.
It concerns the conservation  of  charge condition
$k^\mu J_\mu (k) =0$.
In standard Feynman quantum electrodynamic this condition is not 
satisfied by the individual photon--interaction vertex, but is obtained 
 only by summing over all the different positions where the photon 
interaction can be coupled into a graph.
This feature is the root of many of the difficulties that arise in 
quantum electrodynamics.

Equation (2.9) shows that the conservation -- law property holds for the 
individual {\it quantum} vertex: there is no need to sum over 
different positions.
The classical interaction, on the other hand, has a form that allows one 
easily to sum over all possible locations along a generalized 
propagator, even before multiplication by $k^\mu$.
This summation converts the classical interaction to a sum of two 
interactions, one located at each end of the line associated with the 
generalized propagator. (See, for example, Eq. (4.1) below).
We always perform this summation. Then the classical parts of the interaction 
are shifted to the hard--photon interaction points, at which 
$k^{\mu}J_{\mu}(k)=0$ holds.

 \noindent{\bf 3. Residues of Poles in Generalized Propagators}

Consider a generalized propagator that has only quantum--interaction insertions.
Its general form is, according to (2.9),
$$
\eqalignno{
\prod^{n}_{j=1}&\left[\left( 
\delta_{\mu_j}^{\sigma_j} k^{\rho_j}_j - 
\delta_{\mu_j}^{\rho_j} k^{\sigma_j}_j \right)
\int^\infty_0 d\lambda_j \left(-{\partial\over\partial p^{\rho_j}}
\right)\right]\cr
&( {i\over \sl{p} +\sl{a}-m} 
\gamma_{\sigma_1}
{i\over \sl{p} +\sl{a}+\sl{k}_1-m} 
\gamma_{\sigma_2} 
{i\over\sl{p}+\sl{a}+\sl{k}_1+\sl{k}_2-m}\cr
&\cdots \times \gamma_{\sigma_{n}} 
{i\over \sl{p}+ \sl{a} + \sl{k}_1 \cdots
+ \sl{k}_n -m} \bigg)&(3.1)\cr}
$$
where
$$ 
a=\lambda_1 k_1 + \cdots + \lambda_n k_n .\eqno(3.2)
$$

The singularities of (3.1) that arise from the multiple end--point $\lambda_1
= \lambda_2 = \cdots \lambda_n =0$ lie on the surfaces
$$
p^2_i = m^2,\eqno(3.3)
$$
where
$$
p_i = p + k_1 + k_2 +\cdots +  k_i.\eqno(3.4)
$$
At a point lying on only one of these surfaces the strongest of these
singularities is a pole.

The Feynman function appearing in (3.1) can be decomposed into a sum of
poles times residues.
At the point $a=0$ this gives
$$
\eqalignno{
&{i(\sl{p}+m)\gamma_{\mu_1} i(\sl{p}+\sl{k}_1+m)
\gamma_{\mu_2}\cdots 
\gamma_{\mu_n} i(\sl{p}+\cdots +
\sl{k}_n+m)\over
(p^2-m^2) ((p+k_1)^2-m^2)((p+\cdots + k_n)^2-m^2)}\cr
& \ \ \ \ \ = \sum^n_{i=0} {N_{1i}\over D_{1i}}
{i(\sl{p}_i+m)\over p^2_i-m^2}
{N_{2i}\over D_{2i}},&(3.5)\cr}
$$
where for each $i$ the numerator occurring on the right--hand side of this
equation is identical to the numerator occurring on the left--hand side.
The denominator factors are
$$
D_{1i} = \prod_{j<i} (2p_i k_{ij}+(k_{ij})^2 + i0),\eqno(3.6a)
$$
and
$$
D_{2i} = \prod_{j>i}(2p_i k_{ij} + (k_{ij})^2 + i0),\eqno(3.6b)
$$
where
$$
\ k_{ij} = \sigma_{ij} [(k_1 + \cdots + k_j)-(k_1 + \cdots + k_i)].\eqno(3.7)
$$

     The sign $\sigma_{ij}=\pm$ in (3.7) is specified in reference 14, where it
 is also shown that that the dominant singularity
on  $p^2_i - m^2 =0$ is the function obtained by simply making the replacement
$$                                                                             
\int^\infty_0 d\lambda_j \left( - {\partial\over \partial p^{\rho_j}}\right)
\left( O(p\to p+\lambda_j k_j)\right) \to p_{i\rho_j}(p_ik_j)^{-1}.\eqno(3.8)
$$
Each value of $j$ can be treated in this way.
Thus the dominant singularity of the generalized propagator (3.1) on $p_i^2
- m^2=0$ is
$$
\eqalignno{
   \prod^n_{j=1} 
&\left[ \left( \delta_{\mu_j}^{\sigma_j} k_j^{\rho_j} - \delta_{\mu_j}^{\rho_j}
k_j^{\sigma_j}\right)
p_{i\rho_j} (p_i k_j)^{-1}\right]\cr
 &\times {N_{1i} i(\sl{p}_i+m)N_{2i}\over 
D_{1i}(p^2_i-m^2)D_{2i}}.
&(3.9)\cr}
$$
  
The numerator in (3.9) has, in general, a factor
$$
\eqalignno{
&\ \ \ \ \ i(\sl{p}_i- \sl{k}_i+m)\gamma_{\sigma_i}i(\sl{p}_i+m) 
\gamma_{\sigma_{i+1}}i(\sl{p}_i+\sl{k}_{i+1}+m)\cr
   &=i(\sl{p}_i-\sl{k}_i+m)\gamma_{\sigma_i} i((\sl{p}_i+m)i(2p_{i\sigma_{i+1}}
+ \gamma_{\sigma_{i+1}}\sl{k}_{i+1})\cr
&\ \ \ \ \  +i(\sl{p}_i - \sl{k}_i +m) \gamma_{\sigma_i}
\gamma_{\sigma_{i+1}}(p^2_i-m^2)\cr
&=i(2p_{i\sigma_i}-\sl{k}_i\gamma_{\sigma_i})i(\sl{p}+m)i(2p_{i\sigma_{i+1}}
+ \gamma_{\sigma_{i+1}}\sl{k}_{i+1})\cr
&\ \ \ \ \ +i(p^2_i-m^2)\gamma_{\sigma_i}(2p_{i\sigma_{i+1}}+\gamma_{\sigma_{i+1}}
\sl{k}_{i+1})\cr
&\ \ \ \ \ +i(\sl{p}_i-\sl{k}_i+m)\gamma_{\sigma_i}\gamma_{\sigma_{i+1}}(p^2_i-m^2)
&(3.10)\cr}
$$

The last two terms in the last line of this equation have factors 
$p^2_i-m^2$.
Consequently, they do not contribute to the residue of the pole at
$p_i^2-m^2=0$. The  terms in (3.10) with a factor $2p_{i\sigma_{i+1}}$,
taken in conjunction with the factor in (3.9) coming from $j=i+1$, give
a dependence $2p_{i\rho_j} 2p_{i\sigma_j}$.
This dependence upon the indices $\rho_j$ and $\sigma_j$ is symmetric under
interchange of these two indices.
But the other factor in (3.9) is antisymmetric.
Thus this contribution drops out.
The contribution proportional to $p_{i\sigma_i}$ drops out for similar reasons.

Omitting these terms that do not contribute to the residue of the  pole
at $p^2_i - m^2$ one obtains in place of (3.10)
the factor
$$
(-i\sl{k}_i\gamma_{\sigma_i})i(\sl{p}_i+m)(i\gamma_{\sigma_{i+1}}\sl{k}_{i+1})
\eqno(3.11)
$$
which is first--order in both $\sl{k}_i$ and $\sl{k}_{i+1}$.
That these ``convergence factors''
actually lead to infrared convergence is shown in references 14 and 15.

\noindent{\bf 4. Inclusion of the Classical Interactions}

The arguments of the preceeding section dealt with processes
containing only $Q$--type interactions. In that analysis the order in which 
these $Q$--type interactions were inserted on the line $L$ of $G$ was held 
fixed: each such ordering was considered separately.

In this section the effects of adding $C$--type interaction are considered.
Each $C$--type interactions introduces a coupling $k^\sigma\gamma_\sigma
=\sl{k}$.
Consequently, the Ward identities, illustrated in (2.2),
can be used to simplify the calculation, but only if the contributions
from all orders of its insertion are treated together.
This we shall do.
Thus for $C$--type interactions it is the  operator $\ha$
defined in (2.5) that is to be used rather than the 
operator $\wt$ defined in (2.7).

Consider, then, the generalized propagator obtained by inserting on some line
$L$ of $G$ a set of $n$ interactions of $Q$--type, placed in some definite
order, and a set of $N$ $C$--type interactions, inserted in all orders.
The meromorphic part of the function obtained after the action of the
$n$ operators $\wwt_j$ is given by (3.9).
The action upon this of the $N$ operators $\ha_j$ of (2.5) is obtained by
arguments similar to those that gave (3.9), but differing by  the fact
that (2.5) acts upon the propagator present {\it before\/} the action of $\ha_j$,
and the fact that now both limits of integration contribute, thus giving for
each $\ha_j$ two terms on the right--hand side rather than one.
Thus the action of $N$ such $\ha_j$'s gives $2^N$ terms:
$$
\eqalignno{
\Bigg[ 
   \prod^{n+N}_{j=n+1}   &\ha_{\mu_j}(k_j)
   P_{\mu_1\cdots \mu_n} 
   (p; Q, k_1, Q, k_2, \cdots Q, k_n)
   {\Bigg]}_{Mero}\cr
&= \sum^{2^N}_{\Theta=1} 
  S{gn}(\Theta )\sum^n_{i=0}
  \prod^{n+N}_{j=n+1}
  \left( 
  {ip^\Theta_{i \mu_j}\over p^\Theta_i k_j}
  \right)\cr
&\times 
  \left\{ \prod^n_{j=1}
  \left[
  \left(
  \delta_{\mu j}^{\sigma_j}k_j^{\rho_j}
  - \delta^{\rho_j}_{\mu_j}k_j^{\sigma_j}
  \right)
  \left(
  {p^\Theta_{i\rho_j}\over p^\Theta_ik_j}
  \right)
  \right]
  \right\}
  \cr
&\times 
  {N^\Theta_{1i}\over D^\Theta_{1i}}
  {i(\sl{p}^\Theta_i+m)\over (p^\Theta_i)^2-m^2} 
  {N^\Theta_{2i}\over D^\Theta_{2i}},
&(4.1)
\cr}
$$                                                    
where 
$$
\eqalignno{
\Theta &= (\Theta_{n+1}, \cdots , \
\Theta_{n+N}),\cr
\Theta_j &= +1 \  {\hbox{or}} \ 0,\cr
S{gn}(\Theta )&= (-1)^{\Theta_{n+1}}(-1)^{\Theta_{n+2}}\cdots
(-1)^{\Theta_{n+N}}\cr
p_i^\Theta &= p_i + \Theta_{n+1}k_{n+1} +\cdots + \Theta_{n+N}k_{n+N},\cr
p_i &= p+k_1 + \cdots + k_i,
&(4.2)
\cr}
$$
and the superscript $\Theta$ on the $N$'s and $D$'s means that 
the argument $p_i$
appearing in (3.5) and (3.6) is replaced by $p_i^\Theta$.
Note that even though the action of $\ha_j$ and $\wwt_j$
involve integrations over $\lambda$ and differentiations, the meromorphic
parts of the resulting generalized propagators are expressed by (4.1) in
relatively simple closed form.
These meromorphic parts turn out to give the dominant contributions in the
mesoscopic regime.

The essential simplification obtained by summing over  all orders of the $C$--type
insertions is that after this summation each $C$--type interaction gives
just two terms. 
The first term is just the function before the action 
of $\ha_j$ multiplied by $ip_{i\mu_j} (p_i k_j)^{-1}$; the second is minus
the same thing with $p_i$ replaced by $p_i +k_j$.
Thus, apart from this simple factor, and, for one term, the overall shift
in $p_i$, the function is just the same as it was before the action of
$\ha_j$.
Consequently, the power--counting arguments used for $Q$--type couplings go 
through essentially unchanged. Details can be found in references 14 and 15.

\vskip 9pt
\noindent{\bf 5. Comparison to Other Recent Works}
\vskip 9pt

The problem of formulating quantum electrodynamics in an axiomatic
field-theoretic framework has been examined by Fr\"{o}hlich, Morchio, and
Strocchi$^{8}$ and by D. Buchholz$^9$, with special attention  to the
non-local aspects arising from Gauss' law.  Their main conclusion, as it 
relates to the present work, is that the energy-momentum spectrum 
of the full system can be separated into two parts, the first being the 
photonic asymptotic free-field part, the second being a remainder that: 1) is 
tied to charged particles, 2) is nonlocal relative to the photonic part, and
3) can have a discrete part corresponding to the electron/positron mass. This 
separation is concordant with the structure of the QED 
Hamiltonian, which has a photonic free-field part and an electron/positron
part that incorporates the interaction term $eA^{\mu} J_{\mu}$, but 
no added term corresponding to the non-free part of the electromagnetic field.
It is also in line with the separation of the classical electromagnetic field,  
as derived from the Li\'{e}nard-Wiechert potentials, into a ``velocity'' part 
that is attached (along the light cone) to the moving source particle, and an 
``acceleration'' part that is radiated away. It is the ``velocity''
part, which is tied to the source particle, and which falls off only as 
$r^{-1}$,
that is the origin of the ``nonlocal'' infraparticle structure that
introduces peculiar features into quantum electrodynamics, as compared to 
simple local field theories. 

In the present approach, the quantum analog of 
this entire classical structure is incorporated into the formula for the 
scattering operator by the unitary factor $U(L)$. It was shown in ref. 11, 
Appendix C, that the non-free ``velocity'' 
part of the electromagnetic field generated by 
$U(L)$ contributes in the correct way to the mass of the electrons and
positrons. It gives also the ``Coulomb'' or ``velocity'' part of the
interaction between different charged particles, which is the part of the 
electromagnetic field that gives the main part of Gauss' law asymptotically. 
Thus our formulas supply in a computationally clean way these ``velocity 
field'' contributions that seem so strange when viewed from other points of 
view.
 
Comparisons to the works in references 17 through 22 can be found in
reference 14. 

\noindent{\bf References}

\begin{enumerate}

\item J. Bros {\it in} Mathematical Problems in Theoretical Physics: 
Proc. of the Int. Conf. in Math. Phys. Held in Lausanne Switzerland
Aug 20-25 1979, ed. K. Osterwalder, Lecture Notes in Physics 116, 
Springer-Verlag (1980); 
H. Epstein, V. Glaser, and D. Iagolnitzer, Commun. Math. Phys. {\bf80},
99 (1981).
\item D. Iagolnitzer, {\it Scattering in Quantum Field Theory: The Axiomatic
and Constructive Approaches}, Princeton University Press, Princeton NJ, 
in the series: Princeton Series in Physics. (1993); 
J. Bros, Physica {\bf 124A}, 145 (1984)
\item D. Iagolnitzer and H.P. Stapp, Commun. Math. Phys. {\bf 57}, 1 (1977); 
D. Iagolnitzer, Commun. Math. Phys. {\bf 77}, 251 (1980)
\item T. Kibble, J. Math. Phys. {\bf 9}, 315 (1968);
Phys. Rev. {\bf 173}, 1527 (1968); {\bf 174}, 1883 (1968); {\bf 175}, 1624 (1968).
\item D. Zwanziger, Phys. Rev. {\bf D7}, 1082 (1973).
\item J.K. Storrow, Nuovo Cimento {\bf 54}, 15 (1968).
\item D. Zwanziger, Phys. Rev. {\bf D11}, 3504 (1975); N. Papanicolaou, Ann.
Phys.(N.Y.) {\bf 89}, 425 (1975)
\item J. Fr\"{o}hlich, G. Morchio, and F. Strocchi, Ann.Phys.(N.Y) {\bf 119},
241 (1979); Nucl. Phys. {\bf B211}, 471 (1983); G. Morchio and F. Strocchi,  
{\it in} Fundamental Problems in Gauge Field Theory, eds. G. Velo and 
A.S. Wightman, (NATO ASI Series) Series B:Physics {\bf 141}, 301 (1985).
\item D. Buchholz, Commun. Math. Phys. {\bf 85}, 49 (1982); Phys. Lett. B
{\bf 174}, 331 (1986); {\it in} Fundamental Problems in Gauge Field Theory, eds.
G. Velo and A.S. Wightman, (NATO ASI Series) Series B: Physics {\bf 141}, 381
(1985);
\item T. Kawai and H.P. Stapp, {\it in} 1993 Colloque International en l'honneur de Bernard
Malgrange (Juin, 1993/ at Grenoble)  Annales de l'Institut Fourier {\bf 43.5},
1301 (1993)
\item H.P. Stapp, Phys. Rev. {\bf 28D}, 1386 (1983).
\item F. Block and A. Nordsieck, Phys. Rev. {\bf 52}, 54 (1937).
\item G. Grammer and D.R. Yennie, Phys. Rev. {\bf D8}, 4332 (1973).
\item T. Kawai and H.P. Stapp, Phys. Rev. D {\bf 52}, 2484 (1995).
\item T. Kawai and H.P. Stapp, Phys. Rev. D {\bf 52}, 2505, 2517 (1995).
\item T. Kawai and H.P. Stapp, {\it Quantum Electrodynamics at Large
Distances}, Lawrence Berkeley Laboratory Report LBL-25819 (1993).
\item J. Schwinger Phys. Rev. {\bf 76}, 790 (1949). 
\item D. Yennie, S. Frautschi, and H. Suura, Ann. Phys. (N.Y.) {\bf 13}, 
379 (1961).
\item K.T. Mahanthappa. Phys. Rev. {\bf 126}, 329 (1962); K.T Mahanthappa and 
P.M. Bakshi, J. Math. Phys. {\bf 4}, 1 and 12 (1963).
\item V. Chung, Phys. Rev. {\bf 140}, B1110 (1965)
\item P.P. Kulish and L.D. Fadde'ev, Theor. Math. Phys. {\bf 4}, 745 (1971).
\item E. d'Emilio and M. Mintchev, Fortschr. Phys. {\bf 32}, 473 (1984);
Phys. Rev. {\bf 27}, 1840 (1983)

\end{enumerate}
\vskip .2in
This work was supported by the Director, Office of Energy 
Research, Office of High Energy and Nuclear Physics, Division of High 
Energy Physics of the U.S. Department of Energy under Contract 
DE-AC03-76SF00098.

\end{document}